\documentclass[12pt]{article}

\pdfoutput=1

\usepackage{cite}

\usepackage{amssymb,latexsym}

\usepackage{epstopdf}

\usepackage{graphics,epsfig}

\usepackage{color}

\let\a=\alpha \let\b=\beta \let\g=\gamma \let\d=\delta \let\e=\epsilon
\let\z=\zeta  \let\th=\theta  \let\k=\kappa
\let\l=\lambda \let\m=\mu \let\n=\nu \let\x=\xi \let\p=\pi %\let\r=\rho
\let\s=\sigma   \let\f=\phi  
 
        \let\Th=\Theta \let\L=\Lambda
\let\X=\Xi  \let\S=\Sigma  \let\Y=\Psi
 
\let\la=\label  
  
\def\nn{\nonumber} \def\bd{\begin{document}} \def\ed{\end{document}}
\def\ds{\documentstyle} \let\fr=\frac \let\bl=\bigl \let\br=\bigr
\let\Br=\Bigr \let\Bl=\Bigl
\let\bm=\bibitem
\let\na=\nabla
\def\tU{{\widetilde U}}
\let\pa=\partial \let\ov=\overline
\def\ie{{\it i.e.\ }}
\newcommand{\be}{\begin{equation}}
\newcommand{\ee}{\end{equation}}
\def\ba{\begin{array}}
\def\ea{\end{array}}
\def\ft#1#2{{\textstyle{{\scriptstyle #1}\over {\scriptstyle #2}}}}
\def\fft#1#2{{#1 \over #2}}
\def\F#1#2{{ F_{#1}^{(#2)} }}
\def\cF#1#2{{ {\cal F}_{#1}^{(#2)} }}

\def\R{{\bf R}}
\def\sst#1{{\scriptscriptstyle #1}}
\def\oneone{\rlap 1\mkern4mu{\rm l}}
\def\e7{E_{7(+7)}}
\def\td{\tilde}
\def\wtd{\widetilde}
\def\im{{\rm i}}
\def\bog{Bogomol'nyi\ }
\newcommand{\ho}[1]{$\, ^{#1}$}
\newcommand{\hoch}[1]{$\, ^{#1}$}
\newcommand{\bea}{\begin{eqnarray}}
\newcommand{\eea}{\end{eqnarray}}
\newcommand{\ra}{\rightarrow}
\newcommand{\lra}{\longrightarrow}
\newcommand{\Lra}{\Leftrightarrow}
\newcommand{\ap}{\alpha^\prime}
\newcommand{\bp}{\tilde \beta^\prime}
\newcommand{\cB}{{\cal B}}
\newcommand{\cO}{{\cal O}}
\newcommand{\vecx}{\vec{x}}
\newcommand{\vecy}{\vec{y}}
\newcommand{\vecp}{\vec{p}}
\newcommand{\vecq}{\vec{q}}
\newcommand{\tr}{{\rm tr} }
\newcommand{\Tr}{{\rm Tr} }
\newcommand{\NP}{Nucl. Phys. }

\newcommand{\cL}{{\cal L}}
\newcommand{\cA}{{\cal A}}
\newcommand{\cT}{{\cal T}}
\newcommand{\cD}{{\cal D}}
\newcommand{\cH}{{\cal H}}

\def\sst#1{{\scriptscriptstyle #1}}
\def\0{{\sst{(0)}}}
\def\1{{\sst{(1)}}}
\def\2{{\sst{(2)}}}
\def\3{{\sst{(3)}}}
\def\4{{\sst{(4)}}}
\def\5{{\sst{(5)}}}
\def\6{{\sst{(6)}}}
\def\7{{\sst{(7)}}}
\def\8{{\sst{(8)}}}
\def\9{{\sst{(9)}}}
\def\p{{\sst{(p)}}}
\def\q{{\sst{(q)}}}
\def\ve{\varepsilon}
\def\vf{\varphi}
\def\F{\Phi}
\def\wg{\wedge}

\def\thb{\bar{\theta}}
\def\Thb{\bar{\Theta}}
\def\barp{\bar{p}}
\def\barq{\bar{q}}
\def\barc{\bar{c}}
\def\bard{\bar{d}}
\def\e{\epsilon}

\def \bi{\bibitem}
\def \la {\label}

\def \l {\lambda}
\def\foot{\footnote}
\def \tl  {{\tilde \l}}
\def \sql {{\sqrt \l}}
\def \adss {$AdS_5 \times S^5$\ }
\newcommand{\rf}[1]{(\ref{#1})}
\def \ov {\over}

\def\th{\theta}
\def\Th{\Theta}
\def\vth{\vartheta}
\def\btheta{{\bar\theta}}
\def\ttheta{{{\tilde\theta}}}
\def\bttheta{{{\bar\ttheta}}}
\def\vth{\vartheta}

\def\ra{\rightarrow}
\def\N{\nabla}
\def\F{{\cal F}}
\def\uM{\underline{M}}
\def\uA{\underline{A}}
\def\uN{\underline{N}}
\def\uP{\underline{P}}
\def\ua{\underline{a}}
\def\ub{\underline{b}}
\def\uc{\underline{c}}
\def\ud{\underline{d}}
\def\ue{\underline{e}}
\def\uf{\underline{f}}
\def\ui{\underline{i}}
\def\uj{\underline{j}}
\def\uk{\underline{k}}
\def\ul{\underline{l}}
\def\ual{\underline{\alpha}}
\def\ube{\underline{\beta}}
\def\um{\underline{m}}
\def\un{\underline{n}}
\def\up{\underline{p}}
\def\uq{\underline{q}}
\def\ur{\underline{r}}
\def\us{\underline{s}}
\def\umu{\underline{\mu}}
\def\unu{\underline{\nu}}
\def\ula{\underline{\l}}
\def\uka{\underline{\k}}
\def\usi{\underline{\s}}
\def\urh{\underline{\r}}
\def\cc{\circ}
\def\eqv{\equiv}

\def\ni{\noindent}

\def\Ep{E^{{}^{(+)}}}
\def\Em{E^{{}^{(-)}}}

\def\Mp{M^{{}^{(+)}}}
\def\Mm{M^{{}^{(-)}}}

\def \ha{{1\ov 2}}

\def\r{\rho}

\def\Y{{\rm Y}}
\def\X{{\rm X}}
\def\tY{\tilde{\rm Y}}
\def\tX{\tilde{\rm X}}
\def\dY{\dot{\rm Y}}
\def\dX{\dot{\rm X}}

\def \J {\mathcal{J}}
\def \del {\partial}

\def\dF{\dot{F}}
\def\dG{\dot{G}}
\def\df{\dot{f}}
\def \E {{\cal E}}
\def \S {{\cal S}}
\def \J {{\cal J}}

\def\ms{\mathcal{S}}
\def\mj{\mathcal{J}}
\def\soj{\fr{\ms}{\mj}}
\def \R {{\bf R}}
\def \om {\omega}
\def \bE {\bar E}
\def \x {{\cal X}}

\def \bi{\bibitem}
\def \la {\label}

\def \l {\lambda}
\def\foot{\footnote}
\def \tl  {{\tilde \l}}
\def \sql {{\sqrt \l}}
\def \adss {$AdS_5 \times S^5$\ }
\def \ov {\over}

\def \varpi {{\rm w}}

\def\thb{\bar{\theta}}
\def\Thb{\bar{\Theta}}
\def\mb{\bar{\m}}
\def\ab{\bar{\a}}
\def\zb{\bar{z}}
\def\psib{\bar{\psi}}
\def\barp{\bar{p}}
\def\barq{\bar{q}}
\def\barc{\bar{c}}
\def\bard{\bar{d}}
\def\e{\epsilon}
\def\wb{\bar{w}}
\def\lb{\bar{\l}}
\def\Jb{\bar{J}}
\def\Nb{\bar{N}}
\def\Zb{\bar{Z}}
\def\pab{\bar{\pa}}

\def\At{\tilde{A}}
\def\Bt{\tilde{B}}
\def\Ct{\tilde{C}}
\def\Dt{\tilde{D}}
\def\Et{\tilde{E}}
\def\Ft{\tilde{F}}
\def\Gt{\tilde{G}}
\def\Ht{\tilde{H}}
\def\Mt{\tilde{M}}
\def\Rt{\tilde{R}}
\def\at{\tilde{a}}
\def\bt{\tilde{b}}
\def\ct{\tilde{c}}
\def\dt{\tilde{d}}
\def\et{\tilde{e}}
\def\ft{\tilde{f}}
\def\htil{\tilde{h}}
\def\gt{\tilde{g}}
\def\mt{\tilde{\mu}}
\def\nt{\tilde{\nu}}
\def\pht{\tilde{\f}}

\def\rht{\tilde{\rho}}

\def\asth{\hat{*}}
\def\phh{\hat{\phi}}

\def\bA{{\bf A}}

\def\ola{\overleftarrow}
\def\ora{\overrightarrow}
\def\alt{\tilde{\a}}

\def\eh{\hat{e}}
\def\eph{\hat{\e}}
\def\ph{\hat{p}}
\def\alh{\hat{\a}}
\def\beh{\hat{\b}}
\def\gah{\hat{\g}}
\def\Fh{\hat{F}}
\def\muh{\hat{\m}}
\def\nuh{\hat{\n}}
\def\thh{\hat{\th}}
\def\rhh{\hat{\r}}
\def\dh{\hat{d}}
\def\ih{\hat{i}}
\def\jh{\hat{j}}
\def\kh{\hat{k}}
\def\deh{\hat{\d}}
\def\wh{\hat{w}}
\def\lah{\hat{\l}}
\def\Ah{\hat{A}}
\def\Ch{\hat{C}}
\def\Omh{\hat{\Omega}}

\def\xh{\hat{x}}

\def\ps{\rlap{\, /}\;\,p }
\def\ks{\rlap{\, /}\;\,k }

\def\gym{g_{YM}}

\def\adot{\dot{a}}
\def\bdot{\dot{b}}
\def\bpa{\bar{\pa}}

\def\pr{\prime}
\def\ssk{\medskip}

\begin{document}

\overfullrule=0pt
\parskip=2pt
\parindent=12pt
\headheight=0in \headsep=0in \topmargin=0in
\oddsidemargin=0in

\vspace{ -3cm}
\thispagestyle{empty}
%\vspace{1cm}
%\begin{flushright}
%Preprint DFPD 01/TH/\\
%hep-th/
%\end{flushright}

 \vspace{0.1cm}

\setcounter{equation}{0}
\setcounter{footnote}{0}
\setcounter{section}{0}

\begin{center}

{\Large\bf ADM reduction of IIB on ${\mathcal H}^{p,q}$ and dS braneworld}

\vskip 0.8cm

 \vspace{.5cm}

E. Hatefi\\

%\vspace*{0.8cm}
{ {
{\it International Centre for Theoretical Physics\\
Strada Costiera 11, Trieste, Italy. \\
ehatefi@ictp.it
}}}

\vspace{0.6cm}
A. J. Nurmagambetov\\
{\it A.I. Akhiezer Institute for Theoretical Physics of NSC KIPT\\
1 Akademicheskaya St., Kharkov, UA 61108, Ukraine\\
{
ajn@kipt.kharkov.ua}
}

%%\vspace{0.8cm}
%{\it Kavli Institute for Theoretical Physics,\\
% Santa Barbara, California, USA \\
%} \vspace{0.3cm}

\vspace{0.5cm}
I. Y. Park
%\footnote{Permanent address:
%  Philander Smith College,
%Little Rock, AR 72223, USA
% }
\\

{\it Center for Quantum Spacetime, Sogang University\\
Shinsu-dong 1, Mapo-gu, 121-742 South Korea \\
}
\vspace{0.1cm}
and

\vspace{0.1cm}
{\it Department of Natural and physical Sciences,
Philander Smith College %\footnote{Home institute}
                               \\
Little Rock, AR 72223, USA \\
inyongpark05@gmail.com
}

\end{center}

 \vspace{0.1cm}

 \begin{abstract}
 %%%%%%%%%%%%%%%%%%%%%%%%%%%%%%%%%
We propose a new Kaluza-Klein reduction scheme based on ADM decomposition.
The scheme has been motivated by AdS/CFT, especially by how the
worldvolume theory should appear from the supergravity side.
We apply the scheme to IIB supergravity reduced on a 5D hyperboloidal $\cH^5$ space, and
show that an (A)dS "braneworld" is be realized after
further reduction to 4D. We comment on applications to cosmology and
black hole physics. In particular, the scheme should provide a proper paradigm
for black hole physics.

\end{abstract}
\newpage

%{\color{blue}Questions/comments by AJN. I'll highlight the addressed points in text too.} {\color{red} This color %is for my adds.}

\section{Introduction}

AdS/CFT belongs to a class of dualities in which
the dualization procedure is not explicitly introduced. The reason for that is simple: implementing the procedure is subtle and difficult. Evidently, however, understanding the procedure must be essential for the first principle derivation of AdS/CFT. There has been progress in this direction (see, e.g., \cite{Nurmagambetov:2011pu}\cite{Hatefi:2012sy} and refs therein), and one of the goals in this work is to further that progress.
While doing so, we report on two conceptual/technical advances: the first is a new Kaluza-Klein reduction procedure based on ADM decomposition and the other is
the IIB realization of a braneworld.

\ssk

One of the distinctions of AdS/CFT type dualities is that the dualization and inverse dualization seem very different at low energy field theory level. We refer to the procedure in which one gets the closed string/gravity degrees of freedom from open string/gauge theory as {\em forward dualization}. The {\em inverse dualization} refers to the reverse procedure. It was proposed in \cite{Park:2007mc} that it should be the quantum/strong coupling effects that must be behind the forward dualization. As commented in
\cite{Park:2008sg} (footnote 15), the inverse dualization must be initiated by a spontaneous symmetry breaking of the supergravity system. (Here we are using the term 'spontaneous symmetry breaking' in a general sense that is associated with expanding an action around a solution.\footnote{The symmetries that are broken and the breaking patterns will not be pursued in this work.})

\ssk

\ssk

In the context of $(A)dS/CFT$ type dualities, it is natural to view the $(A)dS_5$ as a foliation
of $(A)dS_4$ along a direction of $(A)dS_5$. (See the figure.)
 One of the leaves can serve as our braneworld with a certain gravity localization mechanism that
 we discuss later. We focus on the $dS$ case henceforth whenever possible. Figure 1 depicts $dS_5$ as foliation of $dS_4$
 hypersurfaces. One may describe the $dS_5$ through bulk gravity setup. On the other hand, it seems plausible to describe
 the bulk $dS_5$, or at least some aspects of it, through collective dynamics of the hypersurfaces.
 Of course, the existence of these "dual" descriptions must be what is behind gauge/gravity
 correspondence. (The bulk dynamics would include the dynamics
 associated with the "radial" direction which is typically associated with renormalization group flow.
 Therefore, in general, the collective surface dynamics would not cover the entire bulk dynamics. There are various levels of equivalence that the term "duality" describes.
Ideally, the term should be reserved only for the cases where the two theories under consideration are fully equivalent. For example, a canonical transformation can be viewed as a precise duality: it maps to a theory that is fully dual to the original theory.
However, even if the full equivalence is not obvious, the term "duality" is often used in some string theory contexts for the cases in which
the two theories capture substantial aspects each other.)

What procedure could lead to the surface degrees of freedom starting from the bulk
  theory? As anticipated in \cite{Park:2008sg}, it should be a procedure initiated by a spontaneous symmetry
 breaking.
It is also likely that the procedure should involve a certain dimensional reduction scheme, conventional or
unconventional.
Although it should be possible to deduce the hypersurface degrees of freedom through the conventional Kaluza-Klein
reduction (see, e.g., \cite{Lehners:2007xa} for a relatively recent discussion), we will pave our way through an unconventional reduction scheme.
 This procedure of acquiring surface degrees of freedom should be viewed as a novel
 Kaluza-Klein (KK) reduction - what we call ADM reduction.
What is unusual about this scheme is that the reduced lagrangian is not a gravity system: the {\em dynamical} fields are the  worldvolume (i.e., the selected hypersurface) gauge fields.
 This phenomenon, unusual in the Kaluza-Klein context, must be what triggers
the inverse dualization of the AdS/CFT type dualities.

\ssk

\begin{figure}
\centerline{
\begin{minipage}[b]{11cm}
             \epsfxsize=11cm
              \epsfbox{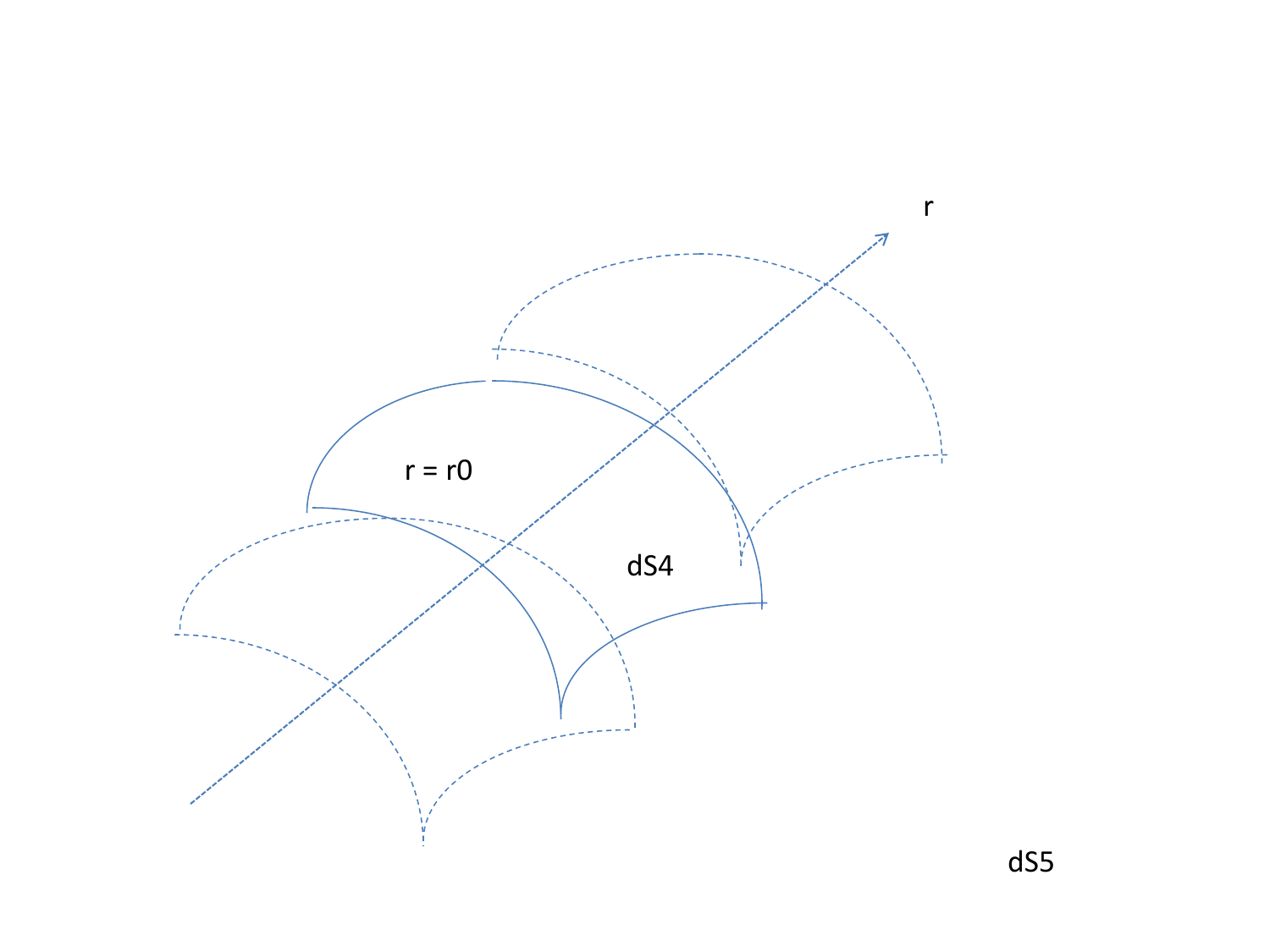}
      \end{minipage}
      }
\caption{$dS_5$ as foliation of $dS_4$}
\label{fig}
\end{figure}

An essential computational ingredient for obtaining the hypersurface degrees of freedom from a spontaneous symmetry breaking was obtained in a remarkable series of papers, \cite{Sato:2002kv,Sato:2003ky,Sato:2004ic}.\footnote{We have withdrawn the claim in earlier versions of this work that the work of \cite{Sato:2002kv} contains an error. The confusion was caused in part by an unusual phenomenon that we discuss in footnote 4.
In short, the ansatz of \cite{Sato:2002kv} used in string frame also leads, as we show below, to consistent reduction in Einstein frame even without the usual dilaton re-scaling.} The authors showed that the Hamilton-Jacobi equation of the gravity system \cite{Parry:1993mw}\cite{Darian:1997mp}\cite{Fukuma:2000bz} under consideration admits a solution of the worldvolume theory form.
We will apply the technique of \cite{Sato:2002kv,Sato:2003ky,Sato:2004ic,Shiromizu:2003dr}
to a specific setup of the 5D gravity that can be obtained by reducing IIB supergravity on a 5D hyperboloidal space $\cH^{p,q}, p+q=5$ considered in \cite{Cvetic:2004km} (and also on $S^5$).

Once the 5D (A)dS gravity is obtained by reducing IIB supergravity on a 5D hyperboloidal space $\cH^5$ ($S_5$) \cite{Cvetic:2004km}, a canonical
transformation can be performed on the system to convert it into an equivalent, therefore dual,
formulation that still takes the form of a supergravity. Following \cite{Sato:2002kv,Sato:2003ky,Sato:2004ic}, one can show that the dual system admits, in the case of $S^5$ reduction, a worldvolume action as solution of the
Hamilton-Jacobi equation of the gravity system.
Obviously, the resulting worldvolume action will capture some aspects of the original gravity
theory, and can be viewed as a dual pair at least in a wider sense of the term mentioned previously.
In essence, the holographic dualities must have their roots in that one may adopt two
different approaches to describe the geometry. In the first approach, one can adopt the
conventional degrees of freedom, the metric, to describe the bulk physics.
In the other approach, one slices the bulk into a set of hypersurfaces, i.e., focus on
each leaf of the foliated geometry.
Gravity is not needed as {\em dynamical} degrees of freedom to describe the
hypersurface, although it serves as a background for the gauge degrees of freedom.
We will elaborate on this in the main body of this paper.

\ssk

Below we consider both the ADM reduction and the standard toroidal-type reduction.
 In the ADM reduction scheme, one employs
the HJ procedure, and a worldvolume effective action will appear as a solution of the HJ equation.
 The ADM reduction scheme can be viewed as "emergent gauge theory" in the sense that a worldvolume gauge field emerges from the symmetry breaking.
In the standard toroidal reduction, dependence of one of the coordinates ("$r$") will be
removed.
The model that we obtain below has a scalar that can be viewed as an inflaton field
in four dimensions. We comment on the potential phenomenological value of
our model in the main body postponing the full analysis for the near future.
In the related literature, usually, an explicit coupling between gravity and various brane sources is employed followed by Calabi-Yau compactification in order to obtain a de Sitter space in the lower dimensions. One drawback of the Calabi-Yau compactification is the implicit nature of the analysis involved. Moreover, the original motivation for considering the Calabi-Yau manifold as opposed to a maximally symmetric manifold has diminished with better understanding of
supersymmetry breaking effects of D-branes.

\vspace{.3in}

The organization of the paper is as follows:
in sec 2, we carry out reduction of IIB supergravity on a manifold denoted by ${\cal M}^5$ that
we take either ${\cal M}^5=\cH^5$ or ${\cal M}^5=S^5$ (5-sphere) in the subsequent sections.
For ${\cal M}=\cH^5$, we obtain a 5D de Sitter
gravity.
  The ADM reduction of the 5D theory obtained thereby is carried out in sec 3.
   In particular, we elaborate on the appearance of the worldvolume gauge field.
 We discuss the implications of our results on braneworld realization and black hole information physics.
 In sec 4, we obtain a domain-wall solution for the 5D system obtained in sec 2 in the case ${\cal M}^5=S^5$.
 Keeping the minimal set of fields, the system admits an AdS vacuum solution. We comment on possibility of obtaining a dS solution with the form fields added.
In another direction, we carry out toroidal reduction to 4D, and obtain an action that may have
phenomenological value for inflationary physics.
In the conclusion, the results are summarized and future directions are suggested.
We also comment on the potential cosmological/black hole applications of our results.

%%%%%%%%%%%%%%%%%%%%%%%%%%%%%%%%%%%%%%%%%%%%%%%%%%%%%%%%%%%%%%%%
\section{Spherical/hyperboloidal reduction in Einstein frame to 5D \la{hr}}
%%%%%%%%%%%%%%%%%%%%%%%%%%%%%%%%%%%%%%%%%%%%%%%%%%%%%%%%%%%%%%%%
\renewcommand{\theequation}{2.\arabic{equation}}
\setcounter{equation}{0}

Although mathematically elegant, the usual Calabi-Yau compactification has one shortcoming:
the requirement of the internal manifolds to be of CY is not sufficiently restrictive, an aspect that
can be seen from the fact that there exists many moduli. Starting with simple compactification such
as compactification on a maximally symmetric space could be more effective.
In the KKLT \cite{Kachru:2003aw} type compactification, one introduces and/or
adds (anti)-branes to lift up the moduli. This can be viewed as a narrowing-down to special
sectors of the moduli space. Therefore, this approach ultimately might not be more general
than the present approach where one restricts to a certain class of special
internal manifolds from the beginning.

In this section, we consider Kaluza-Klein reduction on an inhomogeneous\footnote{Here inhomogeneity refers to the fact that the space is not a coset manifold.} hyperboloidal space $\cH^{p,q}$, a manifold considered in \cite{Cvetic:2004km}.
The authors of \cite{Cvetic:2004km} showed that reduction on $\cH^{p,q}$ leads to a ghost-free dS gravity in the lower dimensions. The ansatz of \cite{Cvetic:2004km} led to a 5D potential that has a saddle shape.
In the reduction that we carry out in this section,
we consider an ansatze that is an analogue of those of \cite{Sato:2002kv}
keeping three scalars
for 5D system: the dilaton, the axion and a breathing mode from the 10D metric. Only the breathing mode
generates the potential for 5D theory as we will see below.\footnote{ The authors of \cite{Sato:2002kv} considered IIB action in string frame.
 The Einstein frame ansatze that we consider below are not connected to the ansatze  considered in \cite{Sato:2002kv}, therefore should belong to a different class of ansatze. ($S_5$ vs $\cH_5$ does not matter for this matter.)
As a matter of fact, one can show that the precise forms of (2.4), (2.5) and (2.6) of
 \cite{Sato:2002kv} used in {\em Einstein} frame also lead to consistent reduction.
 (The precise forms of (2.4-6) of \cite{Sato:2002kv} and our ansatze \rf{ans1q0}, \rf{ans2q0} lead to the same 5D action up to a numerical rescaling, namely, \rf{5dactRhoEinq}.)
 This seems rather unusual, and must be attributed to the simplicity of the ansatze.
 (Our earlier false accusations of the work of \cite{Sato:2002kv} were made in part by this subtlety.)
 In general,
 if an ansatz leads to consistent reduction in one frame, it would not in the other frame without proper dilaton re-scaling of the metric and the other fields.
In one of the footnotes in section 3, we point out another related aspect of the two ansatze.
}

\vspace{.2cm}

The bosonic part of type IIB supergravity action takes the following form in Einstein frame
\cite{DallAgata:1997ju}
%%%
\[
I= \fr{1}{2\k_{10}^2}\int d^{10}X \sqrt{-G_E}\Big[\left( R-\fr12 \pa_M \Phi \pa^M\Phi
  -\fr1{2\cdot 3!}e^{-\Phi}(H_\3)^2 \right)
\]
\[
  -\fr12 e^{2\Phi}\pa_M C_\0 \, \pa^M C_\0 -\fr1{2\cdot 3!} e^{\Phi}(\Ft_\3)^2 -\fr1{4\cdot 5!}(\Gt_\5)^2+\mathcal{L}_{PST}
    \Big]
\]
\be
-\fr{1}{4\k_{10}^2}\int_{\mathcal{M}^{10}}  C_\4\wedge H_\3\wedge F_\3
\la{IIBLagEq0}
\ee
%%%
Let us consider the following reduction ansatz\footnote{In a typical Kaluza-Klein reduction, it is
usually a scalar sector that makes the procedure complicated. An example of sphere reduction with many scalars can be found in \cite{Cvetic:1999xp}.},
%%%
\bea
ds_{10}^2 &=& e^{2\tilde{\r }(\xh)}h_{\um \un}(\xh)d\xh^{\um} d\xh^{\un}+e^{-6\tilde{\r}(\xh)/5}d\Omega_5\equiv e^{2\tilde{\r }(\xh)} h_{\um \un}(\xh)d\xh^{\um} d\xh^{\un}+e^{-6\tilde{\r}(\xh)/5}g_{ij}dy^i dy^j \nn\\
 \Phi (X)&=& \f(\xh)  \nn\\
 B_\2 (X)&=& \fr12 B_{\um \un}(\xh)d\xh^{\um} \wedge d\xh^{\un} \equiv B_\2 (\xh) \nn\\
 C_\0 (X)&=& \chi(\xh) \nn\\
 C_\2 (X)&=& \fr12 C_{\um \un}(\xh)d\xh^{\um} \wedge d\xh^{\un} \equiv C_\2 (\xh)
 \la{ans1q0}
\eea
%%%
and
%%%
\bea
C_\4 (X) &=& \fr{1}{4!}D_{\um \un \uk \ul}(\xh)d\xh^{\um} \wedge d\xh^{\un} \wedge d\xh^{\uk} \wedge d\xh^{\ul}
   +\fr1{4!}k E_{i_1i_2i_3i_4} dy^{i_1} \wedge dy^{i_2} \wedge dy^{i_3} \wedge dy^{i_4}  \nn\\
  &\equiv &  D_\4(\xh)+kE_\4   (y)  \la{ans2q0}
\eea
%%%
The $y$-coordinates describe either $\cH^5$ or $S^5$:
%%%
\bea
{\cal M}^5= \cH^5\quad \mbox{or} \;\;\;S^5
\eea
%%%
 ${\cal M}^5$ is later decided to be $S^5$ for the discussion in sec 3.
The field $E$ satisfies $5 \pa_{[i_1}E_{i_2i_3i_4i_5]}=(1/\sqrt{g})\e_{i_1i_2i_3i_4i_5}$.
(Our form conventions are summarized in Appendix A.)

Substituting \rf{ans1q0}, \rf{ans2q0} into 10D equation of motion, one can show after some algebra that
the reduced field equations follow from the following 5D action:
\[
I= \fr{1}{2\k_{5}^2}\int d^{5}\xh \,\sqrt{-h}\Big[\left( R^\5-\fr{24}{5} \pa_{\um} \tilde{\r} \pa^{\um}\tilde{\r}-\fr12 \pa_{\um}\phi \pa^{\um}\phi
  -\fr1{2\cdot 3!}e^{-\phi-4\tilde{\r}}(H_\3)^2 \right)
\]
\be
  -\fr12 e^{2\phi}\pa_{\um} \chi \, \pa^{\um} \chi -\fr1{2\cdot 3!} e^{\phi-4\tilde{\r}}(\Ft_\3)^2 -\fr1{2\cdot 5!}e^{-8\tilde{\r}}(\Gt_\5)^2+e^{16\tilde{\r}/5}R^{{\cal M}_5}
    \Big]\,
\la{5DLagEins}
\ee
%%%
Above
%%%
\be
\Ft_\3=F_\3-\chi H_\3\equiv dC_\2 (\xh)-\chi(\xh) \wg dB_\2(\xh) \,.
\ee
%%%
\bea
 \Gt_\5 &=& G_\5-C_\2\wedge H_\3,\quad H=\fr12 \pa_{[\um}B_{\un\uk]}d\xh^{\um} \wedge d\xh^{\un} \wedge d\xh^{\uk}\nn\\
 G_\5 &=& \fr1{4!} \pa_{[\um_1}D_{\um_2\cdots \um_5]}d\xh^{\um_1} \wedge \cdots \wedge d\xh^{\um_5} \,.
\eea
%%%
With rescaling
\be
\tilde{\r}=-\fr14 \sqrt{\fr53}\r
\la{rho5d}
\ee
the action \rf{5DLagEins} becomes the following form with canonical kinetic terms:
\[
I= \fr{1}{2\k_{5}^2}\int d^{5}\xh\, \sqrt{-h}\Big[\left( R^\5-\fr12 \pa_{\um} {\r} \pa^{\um}{\r}-\fr12 \pa_{\um}\phi \pa^{\um}\phi
  -\fr1{2\cdot 3!}e^{-\phi+\sqrt{\fr53}{\r}}(H_\3)^2 \right)
\]
\be
  -\fr12 e^{2\phi}\pa_{\um} \chi \, \pa^{\um} \chi -\fr1{2\cdot 3!} e^{\phi+\sqrt{\fr53}{\r}}(\Ft_\3)^2 -\fr1{2\cdot 5!}e^{2\sqrt{\fr53}{\r}}(\Gt_\5)^2+e^{-\sqrt{\fr{16}{15}} \r}R^{{\cal M}_5}
    \Big]
\la{5DLagEins1}
\ee
%%%
Rescaling the $\r$ field further in \rf{5DLagEins1}
\be
\r \ra \sqrt{\fr53}\r
\ee
we arrive at an alternative form of the action:
%%%
\bea
I_5
   &=& \fr1{2\k_5^2}\int d^5\xi \sqrt{-h}\Big[ R^\5-\fr56(\pa\r)^2
-\fr12(\pa \f)^2-\fr1{2\cdot 3!}  e^{-\phi}e^{\fr53 \r} H_\3^2\nn\\
    &&  -\fr12 e^{2\phi}(\pa\chi)^2-\fr1{2\cdot 3!}e^{\phi}e^{\fr53 \r}\Ft^{2}_\3-\fr1{2\cdot 5!}e^{\fr{10}3 \r}\Gt^2_\5+e^{-\fr43\r} R^{{\cal M}_5}\Big]
    \label{5dactRhoEinq}
\eea
%%%

%%%%%%%%%%%%%%%%%%%%%%%%%%%%%%%%%%%%%%%%%%%%%%%%%%%
\section{ADM Reduction from 5D to 4D   \la{ADMred}}
%%%%%%%%%%%%%%%%%%%%%%%%%%%%%%%%%%%%%%%%%%%%%%%%%%%

\renewcommand{\theequation}{3.\arabic{equation}}
\setcounter{equation}{0}

In section 4, we will obtain a solution of a certain three brane configuration.
As a matter of fact, the 5D system admits a whole class of D3-brane solutions as we will see. The first step is to obtain the Hamilton-Jacobi (HJ) equation pertaining to \rf{fstotq}
through a series of manipulations following the works of \cite{Sato:2002kv,Sato:2003ky,Sato:2004ic}. The fact that a class of
 solutions of the HJ system of \rf{fstotq} takes a form of a DBI action has deep
 physical implications. For example, the steps for obtaining the worldvolume form solution to the HJ equation should be viewed as a realization of a reduction scheme.
The reason is that the field equations that follow from the worldvolume action can be viewed as outcome of
 substituting an appropriately constructed Kaluza-Klein gravity ansatze into the 5D Hamilton-Jacobi equation. Therefore the whole procedure is in the usual spirit of Kaluza-Klein reduction; hence it can legitimately be called an ADM "reduction" scheme. Also the resulting D3 action should be
 a dual description, at least in the wider sense of the term "dual".

\subsection{Converting to 5D "string" type frame  \la{5Dsf}}

In the next section, we consider the HJ equation of the 5D gravity system obtained in the previous section. It turns out
more convenient for the purpose at hand to cast \rf{5dactRhoEinq} into another frame which we call 5D "string" frame.
To that end, let us consider
%%%
\bea
(h_{Ein})_{\um\un}=e^{-\fr12 \f+\fr{5}{6} \r}(h_{str})_{\um\un}\,;
\eea
%%%
With this, \rf{5dactRhoEinq} now takes\footnote{In the earlier version of this work, the cross term
$\fr52 \pa_{\um} \f \pa^{\um} \r$ was missed. (This has been pointed by Sato and Tsuchiya.)  }
%%%
\bea
I&=& \fr{1}{2\k_{5}^2}\int d^{5}\xh\, \sqrt{-h}\Big[e^{-\fr34 \f+\fr{{5}}{4}\r}\Big( R^\5-{\fr{{10}}{3}}\N^2\r-\fr{35}{12} \pa_{\um} {\r} \pa^{\um}{\r}+2\N^2\f
 \nn\\
 && -\fr54 \pa_{\um}\phi \pa^{\um}\phi+\fr52 \pa_{\um} \f \pa^{\um }\r-\fr1{2\cdot 3!}(H_\3)^2 \Big)
  -\fr12 e^{\fr54\phi+\fr{{5}}{4}\r}\bigg(\pa_{\um} \chi \, \pa^{\um} \chi +\fr1{ 3!} \Ft_\3^2 +\fr1{ 5!}\Gt_\5^2\bigg)\nn\\
  &&\hspace{2in}+e^{-\fr54 \f+\fr34 \r}R^{{\cal M}_5}
    \Big]
\la{5DLagst2q}
\eea
%%%
After partial integration, one finds
%%%
\[
I= \fr{1}{2\k_{5}^2}\int d^{5}\xh\, \sqrt{-h}\Big[e^{-\fr34 \f+\fr{{5}}{4}\r}\left( R^\5+\fr{5}{4} \pa_{\um} {\r} \pa^{\um}{\r}+\fr14 \pa_{\um}\phi \pa^{\um}\phi
  -\fr52 \pa_{\um}\r \pa^{\um}\phi-\fr1{2\cdot 3!}H_\3^2 \right)
\]
\be
  -\fr12 e^{\fr54\phi+\fr{{5}}{4}\r}\bigg(\pa_{\um} \chi \, \pa^{\um} \chi +\fr1{ 3!} \Ft_\3^2 +\fr1{ 5!}\Gt_\5^2\bigg)+e^{-\fr54 \f+\fr34 \r}R^{{\cal M}_5}
    \Big]
\la{5DLagst2q}
\ee
%%%
Let us split the index $\um$:
%%%
\bea
\um=\m,r
\eea
%%%
Carrying out ADM decomposition and adding GH boundary terms yields (see, e.g.,\cite{Embacher:1995xp})
%%%
\bea
%\int d^5x\cL_{total}=
\int\, d^5\xh\,\cL_{bulk+bd}
&=&\int \, dr d^4 x\,\sqrt{-g}\, n
 \left[\fr{}{}\right. e^{-\fr34 \f+\fr{{5}}{4}\r}
 \bigg(
-K_{\mu\nu}^2+K^2\nn\\
&& -\fr{3}{2n}[\pa_r \phi-n^{\mu}\pa_{\mu}\phi]K +\fr{5}{2n}[\pa_r \r-n^{\mu}\pa_{\mu}\r]K \nn\\
&&+\frac{1}{n^2}\bigg\{\frac{1}{4}[\pa_r \phi-n^{\mu}\pa_{\mu}\phi]^2
   +\frac{5}{4}[\pa_r \r-n^{\mu}\pa_{\mu}\r]^2\nn\\
&&-\fr52[\pa_r \phi-n^{\mu}\pa_{\mu}\phi][\pa_r \r-n^{\n}\pa_{\n}\r]
-\frac{1}{4}[H_{r\mu\nu}-n^{\lambda}H_{\lambda\mu\nu}]^2
   \bigg\}\bigg)\nn\\
&&-\frac{1}{2n^2}e^{\fr54\phi+\fr{{5}}{4}\r}\bigg\{[\pa_r \chi-n^{\mu}\pa_{\mu}\chi]^2
+\frac{1}{2}[\tilde{F}_{r\mu\nu}-n^{\lambda}\tilde{F}_{\lambda\mu\nu}]^2
\nn\\
&&+\frac{1}{24}[\tilde{G}_{r\mu\nu\lambda\rho}
-n^{\sigma}\tilde{G}_{\sigma\mu\nu\lambda\rho}]^2 \bigg\} +{\cal L}^\4 \;\;\bigg],
\label{fstotq}
\eea
%%%
where $r$ is one of the spatial coordinates and will play the role of "time" in the next section, and
%%%
\bea
{\cal L}^\4 &\equiv & e^{-\fr34\f+\fr54 \r}\Big(R^\4 +\fr32 \N_\m\N^\m \f-\fr52 \N_\m\N^\m \r
-\fr78\pa_\m \f \pa^\m \f -\fr{15}{8}\pa_\m \r\pa^\m \r+\fr54 \pa_\m \f \pa^\m \r
-\fr1{12}H_{\m\n\r} H^{\m\n\r}\Big)  \nn\\
          &&  -\fr12 e^{\fr54\phi+\fr{{5}}{4}\r}\bigg(\pa_{\m} \chi \, \pa^{\m} \chi +\fr1{ 3!} \Ft^{\m\n\r}\Ft_{\m\n\r} \bigg)+e^{-\fr54 \f+\fr34 \r}R^{{\cal M}_5}
\eea
%%%
With $r$ playing the role of "time", the total "hamiltonian" of the system\footnote{We choose the Hamiltonian as follows:
\[
S=\int \, d^5 \xh \,{\cL}_{bulk+bd}=\int \, dr d^4 x \, \sqrt{-g}\left(\bf{P}\cdot \pa_r{\bf Q}-\cH \right)\,,
\]
where $\bf{Q}$ and $\bf{P}$ are the ``coordinates'' and the corresponding ``momenta''
\[
{\bf{P}}=\fr1{\sqrt{-g}}\fr{\d S}{\d \pa_r \bf{Q}} \, .
\]
.
}
%%%
\bea
\cH&=& \pi^{\m\n}\pa_r g_{\m\n}+\pi_{\phi}\pa_r \phi
+\pi_{\rho}\pa_r \rho+\pi_B^{\mu\nu} \pa_r B_{\mu\nu}
+\pi_{\chi}\pa_r \chi
+\pi_C^{\mu\nu} \pa_r C_{\mu\nu}
+\pi_D^{\mu\nu\lambda\rho} \pa_r D_{\mu\nu\lambda\rho}\nn\\
&&\hspace{3in} -\cL_{bulk+bd}
  \la{pqdq}
\eea
%%%
can be written as (see, e.g., \cite{Sato:2002kv})
%%%
\bea
\cH\equiv nH+n_{\mu}H^{\mu}+B_{r\mu}Z_B^{\mu}+C_{r\mu}Z_C^{\mu}
+D_{r\mu\nu\lambda}Z_D^{\mu\nu\lambda}  \la{hamildefq}
\eea
%%%
Here $n,\; n_{\mu}, \; B_{r\mu}$, $C_{r\mu}$ and $D_{r\mu\nu\lambda}$ behave
like Lagrange multipliers, giving the following set of constraints
\bea
H=0, \;\;\; H^{\mu}=0, \;\;\; Z_B^{\mu}=0, \;\;\;
Z_C^{\mu}=0, \;\;\; Z_D^{\mu\nu\lambda}=0.
\label{constraintsq}
\eea
%%%
These constraints will play an important role in the emergence of the worldvolume gauge field
as we discuss in section 3.3.
One can show
%%%
\bea
H&=& e^{\fr34 \f-\fr54 \r}\bigg( -\pi_{\m\n}^2-\fr12\pi_\m^\m \pi_\f
    +\fr12\pi_\m^\m \pi_\r-\fr12\pi_\f^2-\fr{3}{10}\pi_\r^2  \nn\\
  &&  \hspace{2in}-(\pi_{B\m\n}+\chi \pi_{C\m\n}+6C^{\l\r}\pi_{D\m\n\l\r})^2       \bigg)\nn\\
  &&- e^{-\fr54 \f-\fr54 \r}\bigg(\fr12 \pi_\chi^2+\pi_{C\m\n}^2+12\pi_{D\m\n\l\r}^2
      \bigg)-\cL^\4
\eea
%%%
and
%%%
\bea
H^\m &=& -2\N_\n \pi^{\m\n}+\pi_\f \pa^\m \f+\pi_\r \pa^\m \r+\pi_{B \n\l}H^{\m\n\l}   \nn\\
     && +\pi_{\chi}\pa^\m \chi+\pi_{C\n\l}F^{\m\n\l}+ \pi_{D\n\l\r\s}(G^{\m\n\l\r\s}-4C^{\m\n}H^{\l\r\s}) \nn\\
Z_B^\m &=& 2\N_\n \pi_B^{\m\n}  \nn\\
Z_C^\m &=& 2\N_\n \pi_C^{\m\n}+4\pi_D^{\m\n\l\r}H_{\n\l\r} \nn\\
Z_D^{\m\n\l}&=& 4 \N_\r \pi_D^{\m\n\l\r}  \la{HZconstr}
\eea
%%%
Define $\bar{g}_{\mu\nu}(x,r)$ to be a classical solution to the field equation associated with \rf{fstotq} with the following boundary condition\footnote{
The $r=r_0$ surface should not be taken as a genuine boundary. If it were a genuine boundary,
the metric on the boundary would be a constant as implied by the Dirichlet boundary condition. (Recall that the GH boundary terms were introduced for Dirichlet boundary conditions for the metric.) Rather it should be taken as a device that bridges the bulk description and the hypersurface description.
}
%%%
\bea
 \bar{g}_{\mu\nu}(x,r_0)&=&g_{\mu\nu}(x)\quad,\quad
 \pi^{\mu\nu}(x)=\bar{\pi}^{\mu\nu}(x,r=r_0)
\label{bcq}
\eea
%%%
The boundary configurations for other fields are similarly defined.
The standard procedure of the HJ formalism yields,
%%%
\bea
&&\pi^{\mu\nu}(x)=\frac{1}{\sqrt{-g(x)}}
\frac{\delta S}{\delta g_{\mu\nu}(x)},\;\;\;
\label{pianddelSq}
\eea
%%%
and similarly for other fields. The HJ equation reads
%%%
\bea
&&e^{\fr34 \f-\fr54\r} \left[- \bigg(\frac{1}{\sqrt{-g}}\frac{\delta S_0}{\delta g_{\mu\nu}}\bigg)^2
-\fr{1}{2}\frac{g_{\mu\nu}}{\sqrt{-g}}\frac{\delta S_0}{\delta g_{\mu\nu}}
      \frac{1}{\sqrt{-g}}\frac{\delta S_0}{\delta \phi}
-\fr1{2}\bigg(\frac{1}{\sqrt{-g}}\frac{\delta S_0}{\delta \f}\bigg)^2  \right. \nn\\
&&\left. \hspace{.5in}
 -\fr{3}{10}\bigg(\frac{1}{\sqrt{-g}}\frac{\delta S_0}{\delta \r}\bigg)^2
    +\fr1{2}\frac{g_{\mu\nu}}{\sqrt{-g}}\frac{\delta S_0}{\delta g_{\mu\nu}} \frac{1}{\sqrt{-g}}\frac{\delta S_0}{\delta \r}\right. \nn\\
&&\left. \hspace{1.5in}-\frac{1}{(\sqrt{-g})^2}\bigg(\frac{\delta S_0}{\delta B_{\mu\nu}}
+\chi\frac{\delta S_0}{\delta C_{\mu\nu}}
+6C_{\lambda\rho}
\frac{\delta S_0}{\delta D_{\mu\nu\lambda\rho}}\bigg)^2\right]  \nn\\
&&-\fr{ e^{-\fr54\f-\fr54 \r}}{(\sqrt{-g})^2}\left[\bigg(\frac{1}{2}\frac{\delta S_0}{\delta \chi}\bigg)^2
+\bigg(\frac{\delta S_0}{\delta C_{\mu\nu}}\bigg)^2
+12\bigg(\frac{\delta S_0}{\delta D_{\mu\nu\lambda\rho}}
\bigg)^2\right]\nn\\
&&-\Big[e^{-\fr34\f+\fr54 \r}\Big(R^\4 +\fr32 \N_\m\N^\m \f-\fr52 \N_\m\N^\m \r
-\fr78\pa_\m \f \pa^\m \f -\fr{15}{8}\pa_\m \r\pa^\m \r+\fr54 \pa_\m \f \pa^\m \r \Big)  \nn\\
 &&-\fr1{12} H_{\m\n\r}H^{\m\n\r} -\fr12 e^{\fr54\phi+\fr{{5}}{4}\r}\bigg(\pa_{\m} \chi \, \pa^{\m} \chi +\fr1{ 3!} \Ft^{\m\n\r}\Ft_{\m\n\r} \bigg)+e^{-\fr54\f+\frac{3}{4}\r}R^{{\cal M}_5}\Big]\nn\\
 &&=0
\label{HJfullst}
\eea
%%%
The action \rf{HJfullst} does admit a DBI form of \rf{szero} once we
assume that the fields are constant on the fixed "time" surface
as in \cite{Sato:2002kv,Sato:2003ky,Sato:2004ic}. (In that case, only the $R^{{\cal M}_5}$ term contributes
to the HJ equation among the terms in $\cL^\4$.) We will discuss this
in the next subsection. Also, we discuss the HJ procedure keeping all the terms in
$\cL^\4$. An interesting toggle between $S_5$ and $\cH_5$ will be noted depending
on whether one uses constant field approximation.

For the solution, one can perform the type of derivative expansion
considered, e.g., in \cite{Darian:1997mp}\cite{Shiromizu:2003dr}.
Let us set
%%%
\bea
S_0\equiv S_0^\0+S_0^\1+\cdots  \la{sersol}
\eea
%%%
where $S_0^\0$ ($S_0^\1$) represents the leading (next) order term in the derivative
expansion. Let us work out $S_0^\0$ and $S_0^\1$. In the leading order, the HJ equation reads
%%%
\bea
&&\fr{e^{\fr34 \f-\fr54\r}}{(\sqrt{-g})^2} \left[- \bigg(\frac{\delta S_0^\0}{\delta g_{\mu\nu}}\bigg)^2
-\fr{1}{2}{g_{\mu\nu}}\frac{\delta S_0^\0}{\delta g_{\mu\nu}}
      \frac{\delta S_0^\0}{\delta \phi}
-\fr1{2}\bigg(\frac{\delta S_0^\0}{\delta \f}\bigg)^2  \right. \nn\\
&&\left. \qquad\qquad
-\fr{3}{10}\bigg(\frac{\delta S_0^\0}{\delta \r}\bigg)^2
    +\fr1{2}{g_{\mu\nu}}\frac{\delta S_0^\0}{\delta g_{\mu\nu}} \frac{\delta S_0^\0}{\delta \r}\right]  \nn\\
&&-\fr{ e^{-\fr54\f-\fr54 \r}}{(\sqrt{-g})^2}\left[
12\bigg(\frac{\delta S_0^\0}{\delta D_{\mu\nu\lambda\rho}}
\bigg)^2\right]
-e^{-\fr54\f+\frac{3}{4}\r}R^{{\cal M}_5}=0
\label{HJzeroth}
\eea
%%%
Although \rf{szero} is not a solution once the terms with derivatives in $\cL^\4$ are included, the similar types of the terms that would appear when \rf{szero} is expanded
should appear in the solution. Guided by this let us try the following ansatz
%%%
\bea
S_0^\0=\b_\0 \int d^4x\,\sqrt{-g}\; e^{-\f+\r}  \la{Szz}
\eea
%%%
Substituting \rf{Szz} into \rf{HJzeroth} leads to
%%%
\bea
\fr{1}{5}\b_\0^2=R^{{\cal M}_5}
\eea
%%%
This indicates that one should take ${\cal M}_5=S_5$.
At the next order, the HJ equation takes
%%%
\bea
&&\fr{e^{\fr34 \f-\fr54\r}}{(\sqrt{-g})^2} \left[
-2 \frac{\delta S_0^\0}{\delta g_{\mu\nu}}\frac{\delta S_0^\1}{\delta g_{\r\s}}g_{\m\r}g_{\n\s}
-\fr{1}{2}{g_{\mu\nu}}\frac{\delta S_0^\0}{\delta g_{\mu\nu}}
      \frac{\delta S_0^\1}{\delta \phi}
-\fr{1}{2}{g_{\mu\nu}}\frac{\delta S_0^\1}{\delta g_{\mu\nu}}
      \frac{\delta S_0^\0}{\delta \phi} \right. \nn\\
&&\left. \hspace{.5in}-\frac{\delta S_0^\0}{\delta \f}\frac{\delta S_0^\1}{\delta \f}
-\fr{3}{5}\frac{\delta S_0^\0}{\delta \r}\frac{\delta S_0^\1}{\delta \r}
   +\fr1{2}{g_{\mu\nu}}\frac{\delta S_0^\0}{\delta g_{\mu\nu}}\frac{\delta S_0^\1}{\delta \r}
    \right. \nn\\
&&\left. \hspace{.5in}
+\fr1{2}{g_{\mu\nu}}\frac{\delta S_0^\1}{\delta g_{\mu\nu}}\frac{\delta S_0^\0}{\delta \r}
-\bigg(\frac{\delta S_0^\1}{\delta B_{\mu\nu}}
+\chi\frac{\delta S_0^\1}{\delta C_{\mu\nu}}
+6C_{\lambda\rho}
\frac{\delta S_0^\0}{\delta D_{\mu\nu\lambda\rho}}\bigg)^2\right]  \nn\\
&&- \fr{{e^{-\fr54\f-\fr54 \r}}}{(\sqrt{-g})^2}\left[
\bigg(\frac{\delta S_0^\1}{\delta C_{\mu\nu}}\bigg)^2
\right]
-\Big[e^{-\fr34\f+\fr54 \r}\Big(R^\4 +\fr32 \N_\m\N^\m \f-\fr52 \N_\m\N^\m \r
-\fr78\pa_\m \f \pa^\m \f \nn\\
&&-\fr{15}{8}\pa_\m \r\pa^\m \r+\fr54 \pa_\m \f \pa^\m \r  -\fr{1}{12} H_{\m\n\r}H^{\m\n\r}\Big)
   -\fr12 e^{\fr54\phi+\fr{{5}}{4}\r}\bigg(\pa_{\m} \chi \, \pa^{\m} \chi +\fr1{ 3!} \Ft^{\m\n\r}\Ft_{\m\n\r} \bigg)\Big]\nn\\
   &&=0
\label{HJ1st}
\eea
%%%
With the zeroth order solution \rf{Szz} substituted, this reduces to
%%%
\bea
&&\fr{e^{-\fr14 \f-\fr14\r}}{\sqrt{-g}}\b_\0 \left[
      \fr{2}{5} \frac{\delta S_0^\1}{\delta \r}
-\fr{e^{\f-\r}}{\b_\0 \sqrt{-g}}\bigg(\frac{\delta S_0^\1}{\delta B_{\mu\nu}}
+\chi\frac{\delta S_0^\1}{\delta C_{\mu\nu}}
\bigg)^2\right] - \fr{e^{-\fr54\f-\fr54 \r}}{(\sqrt{-g})^2}
\bigg(\frac{\delta S_0^\1}{\delta C_{\mu\nu}}\bigg)^2
 \nn\\
&&
-\Big[e^{-\fr34\f+\fr54 \r}\Big(R^\4 +\fr32 \N_\m\N^\m \f-\fr52 \N_\m\N^\m \r
-\fr78\pa_\m \f \pa^\m \f \nn\\
&&-\fr{15}{8}\pa_\m \r\pa^\m \r+\fr54 \pa_\m \f \pa^\m \r   -\fr{1}{12} H_{\m\n\r}H^{\m\n\r}\Big)
   -\fr12 e^{\fr54\phi+\fr{{5}}{4}\r}\bigg(\pa_{\m} \chi \, \pa^{\m} \chi +\fr1{ 3!} \Ft^{\m\n\r}\Ft_{\m\n\r} \bigg)\Big]\nn\\
  && =0
\label{HJ1stred}
\eea
%%%
One can show that \rf{HJ1stred} admits the following form of the solution,
%%%
\bea
S_0^\1&=& \fr1{\b_\0}\int d^4x \sqrt{-g} e^{-\fr12\f+\fr32\r}
      \Big[\fr53 R^\4+\fr{55}{24} (\pa \f)^2+\fr{25}{8}(\pa \r)^2-\fr52 (\pa \f\cdot \pa \r) \Big]\nn\\
         &&+\fr{5}{6\b_\0} \int d^4x \sqrt{-g}\;e^{-\f+\r} (\pa \chi)^2
           -\fr{1}{10}\b_\0 \int d^4x \sqrt{-g}e^{\fr32\f+\fr32\r} {\cal F}_{\m\n}{\cal F}^{\m\n} 
           \la{s01} \nn\\
\eea
%%%
 It should be possible to find $S_0^{n}, n>1$ in a similar way.

\subsection{Reduction that leads to DBI form solution to the HJ equation  }

 Let us verify that the HJ equation associated with \rf{fstotq}
admits a DBI form solution in the setup of constant supergravity fields.
In order to make our analysis slightly more general, note the following freedom. Suppose we use
%%%
\bea
\pht\ra a\phh+b\rhh \qquad  \rht\ra c\phh +d\rhh
\eea
%%%
in the ansatze \rf{ans1q0}. Then one should get \rf{5DLagEins} where $\pht,\rht$ are
replaced by $a\phh+b\rhh,\phh +d\rhh$ respectively.
Let us utilize this freedom and introduce the following linear combinations of $\f, \r$,
%%%
\bea
\f &\equiv& a\phh+b\rhh \nn\\
\r &\equiv & c\phh+d \rhh
\eea
%%%
Now Eq.\rf{fstotq} can be rewritten
%%%
\bea
%\int d^5x\cL_{total}=
&&\int\, d^5\xh\,\cL_{bulk+bd}\nn\\
=&&\int \, dr d^4 x\,\sqrt{-g}\, n
 \left[\fr{}{}\right. e^{-\fr34 (a\phh+b\rhh)+\fr{{5}}{4}(c\phh+d \rhh)}
 \bigg(
-K_{\mu\nu}^2+K^2 +\fr{u_1}{n}f(\phh)K +\fr{u_2}{n}f(\rhh)K \nn\\
&&+\frac{1}{n^2}\bigg\{u_3f(\phh)^2
   +u_4f(\rhh)^2+u_5f(\phh)f(\rhh)
-\frac{1}{4}[H_{r\mu\nu}-n^{\lambda}H_{\lambda\mu\nu}]^2
   \bigg\}\bigg)\nn\\
&&-\frac{1}{2n^2}e^{\fr54(a\phh+b\rhh)+\fr{{5}}{4}(c\phh+d \rhh)}\bigg\{[\pa_r \chi-n^{\mu}\pa_{\mu}\chi]^2
+\frac{1}{2}[\tilde{F}_{r\mu\nu}-n^{\lambda}\tilde{F}_{\lambda\mu\nu}]^2
\nn\\
&&+\frac{1}{24}[\tilde{G}_{r\mu\nu\lambda\rho}
-n^{\sigma}\tilde{G}_{\sigma\mu\nu\lambda\rho}]^2 \bigg\} +{\cal L}^\4 \;\;\bigg],
\label{fstotq2}
\eea
%%%
where
%%%
\bea
f(\phh)&\equiv & \pa_r \phh-n^{\mu}\pa_{\mu}\phh \nn\\
f(\rhh)&\equiv & \pa_r \rhh-n^{\mu}\pa_{\mu}\rhh
\eea
%%%
and the $u$'s are related to $a,b,c,d$
%%%
\bea
&& u_1\equiv -\fr32 a+\fr52 c, \qquad u_2 \equiv  -\fr32 b+\fr52 d,\qquad
u_3\equiv \fr14 a^2+\fr54 c^2-\fr{5}{2}ac,\nn\\
&&  u_4\equiv \fr14 b^2+\fr54 d^2-\fr{5}{2}bd,\qquad
u_5 \equiv \fr12 ab+\fr52 cd-\fr52 bc-\fr52 ad
\la{us}
\eea
%%%
%The first of these is the hamiltonian constraint which we will focus on shortly.
After some algebra, one can show
%%%
\bea
\!\!\!\!\!\!\!\!\!H&=& e^{\fr34 (a\phh+b\rhh)-\fr54 (c\phh+d\rhh)}\bigg( -\pi_{\m\n}^2+w_1(\pi_\m^\m)^2+w_2\pi_{\rhh}^2+w_3\pi_{\phh}^2
+w_4\pi_\m^\m \pi_{\phh}
    +w_5\pi_\m^\m \pi_{\rhh}+w_6\pi_{\phh} \pi_{\rhh}  \nn\\
  &&  \hspace{2in}-(\pi_{B\m\n}+\chi \pi_{C\m\n}+6C^{\l\r}\pi_{D\m\n\l\r})^2       \bigg)\nn\\
  &&- e^{-\fr54 (a\phh+b\rhh)-\fr54 (c\phh+d\rhh)}\bigg(\fr12 \pi_\chi^2+\pi_{C\m\n}^2+12\pi_{D\m\n\l\r}^2
      \bigg)-\cL^\4
      \la{Hmix}
\eea
%%%
The parameters $w_1,...,w_6$ are related to $u$'s that appear in \rf{us} by
%%%
\bea
w_1 &\equiv& \fr{u_2^2 u_3+u_1^2 u_4-4 u_3 u_4-u_1u_2u_5+u_5^2}{D} \nn\\
w_2 &\equiv & \fr{u_1^2-3u_3}{D}\quad w_3 \equiv  \fr{u_2^2-3u_4}{D}\nn\\
w_4 &\equiv & \fr{2u_1u_4-u_2u_5}{D}\quad  w_5 \equiv  \fr{2u_2u_3-u_1u_5}{D}\quad
w_6 \equiv  -\fr{2u_1u_2-3u_5}{D}
\eea
%%%
with
%%%
\bea
D\equiv 4u_2^2 u_3+4u_1^2 u_4-12 u_3 u_4-4u_1u_2u_5+3u_5^2
\eea
%%%
As in \cite{Sato:2002kv,Sato:2003ky,Sato:2004ic}, we assume that the fields are constant on the fixed "time" surface. Due to this assumption, only the $R^{{\cal M}_5}$ term contributes
to the HJ equation among the terms in $\cL^\4$; substituting \rf{pianddelSq} into the hamiltonian
 constraint, one finds the following HJ equation:
%%%
\bea
&&e^{\fr34 (a\phh+b\rhh)-\fr54(c\phh+d\rhh)} \left[- \bigg(\frac{1}{\sqrt{-g}}\frac{\delta S_0}{\delta g_{\mu\nu}}\bigg)^2+w_1\bigg(\frac{g_{\m\n}}{\sqrt{-g}}\frac{\delta S_0}{\delta g_{\mu\nu}}\bigg)^2
+w_4\frac{g_{\mu\nu}}{\sqrt{-g}}\frac{\delta S_0}{\delta g_{\mu\nu}}
      \frac{1}{\sqrt{-g}}\frac{\delta S_0}{\delta \phh}
  \right. \nn\\
&&\left. +w_3\bigg(\frac{1}{\sqrt{-g}}\frac{\delta S_0}{\delta \phh}\bigg)^2
+w_2\bigg(\frac{1}{\sqrt{-g}}\frac{\delta S_0}{\delta \rhh}\bigg)^2
+w_6 \frac{1}{\sqrt{-g}}\frac{\delta S_0}{\delta \phh}\frac{1}{\sqrt{-g}}\frac{\delta S_0}{\delta \rhh}
  +w_5  \frac{g_{\mu\nu}}{\sqrt{-g}}\frac{\delta S_0}{\delta g_{\mu\nu}} \frac{1}{\sqrt{-g}}\frac{\delta S_0}{\delta \rhh}\right. \nn\\
&&\left. \hspace{1.5in}-\bigg(\frac{1}{\sqrt{-g}}\frac{\delta S_0}{\delta B_{\mu\nu}}
+\chi\frac{1}{\sqrt{-g}}\frac{\delta S_0}{\delta C_{\mu\nu}}
+6C_{\lambda\rho}\frac{1}{\sqrt{-g}}
\frac{\delta S_0}{\delta D_{\mu\nu\lambda\rho}}\bigg)^2\right]  \nn\\
&&- e^{-\fr54(a\phh+b\rhh)-\fr54 (c\phh+d\rhh)}\left[\bigg(\frac{1}{2}\frac{1}{\sqrt{-g}}\frac{\delta S_0}{\delta \chi}\bigg)^2
+\bigg(\frac{1}{\sqrt{-g}}\frac{\delta S_0}{\delta C_{\mu\nu}}\bigg)^2
+12\bigg(\frac{1}{\sqrt{-g}}\frac{\delta S_0}{\delta D_{\mu\nu\lambda\rho}}
\bigg)^2\right]\nn\\
&&  =
e^{-\fr54(a\phh+b\rhh)+\frac{3}{4}(c\phh+d\rhh)}R^{{\cal M}^5}
\label{HJmod}
\eea
%%%
Following \cite{Sato:2002kv}, let us examine whether \rf{HJmod} admits the following form
of the solution, which is a slight modification of the corresponding solution in \cite{Sato:2002kv}:
%%%
\bea
S_0=S_c+S_{DBI}+S_{WZ} \la{szero}
\eea
%%%
where
\be
S_c=\alpha \int d^4x \sqrt{-g}\, e^{\z_3\phh+\z_4\rhh}
\ee
%%%
%%%
\be
S_{DBI}=\beta \int d^4x\, e^{\z_1\phh+\z_2\rhh} \sqrt{-\det (g_{\mu\nu}+{\cal F}_{\mu\nu})}
\ee
%%%
\bea
S_{WZ}
= \gamma \int d^4x \;\epsilon^{\mu\nu\lambda\rho}
\left( \frac{1}{24}D_{\mu\nu\lambda\rho}
+\frac{1}{4}C_{\mu\nu}{\cal F}_{\lambda\rho}
+\frac{1}{8}\chi {\cal F}_{\mu\nu}{\cal F}_{\lambda\rho} \right),
\label{ScSBISWZ66}
\eea
%%%
where
%%%
\bea
\F_{\m\n}\equiv -B_{\m\n}+F_{\m\n}
\eea
%%%
Inspection of the terms' structures reveals that the presence of $(\pi_\m^\m)^2$
would require a major modification of \rf{szero}; let us impose
%%%
\bea
w_1=0 \la{wconstr}
\eea
%%%
Detailed computation implies that \rf{HJmod} would admit a solution of the form \rf{szero}
once the following conditions are imposed in addition to the previous condition \rf{wconstr}:
%%%
\bea
&& 2w_4\z_1+2w_5 \z_2=1,\quad \b^2(w_3\z_1^2+w_2\z_2^2+w_6\z_1\z_2)=-\fr12 \g^2,\nn\\
&& \z_1=-a,\quad \z_2=-b,\quad   \z_3=-a+c,\quad \z_4=-b+d \nn\\
&& \a^2(-1+2w_4 \z_3+w_3 \z_3^2+w_2 \z_4^2+w_6 \z_3\z_4+2w_5 \z_4)=R^{{\cal M}_5},\nn\\
&& w_4\z_3+w_5\z_4=1,\quad
2w_4\z_1+2w_3\z_1\z_3+2w_2\z_2\z_4+w_6\z_2\z_3+w_6\z_1\z_4+2w_5\z_2=0
 \la{wconstr2}\nn\\
\eea
%%%
The constraints \rf{wconstr} and \rf{wconstr2} amount to 4 constraints among $a,b,c$ and $d$.
In other words, one can first use the second line to replace $\z$'s by the corresponding
expressions on the right-hand sides of the second line. One can then solve
%%%
\bea \la{weq}
&& w_1=0,\quad 2w_4\z_1+2w_5 \z_2=1,\quad  w_4\z_3+w_5\z_4=1 \\
&&
2w_4\z_1+2w_3\z_1\z_3+2w_2\z_2\z_4+w_6\z_2\z_3+w_6\z_1\z_4+2w_5\z_2=0  \nn
\eea
%%%
where $\z$'s should take the explicit expression in terms of $(a,b,c,d)$.
Once the solutions are determined, they can be substituted into the remaining two equations
%%%
\bea
&& \b^2(w_3\z_1^2+w_2\z_2^2+w_6\z_1\z_2)=-\fr12 \g^2  \nn\\
&& \a^2(-1+2w_4 \z_3+w_3 \z_3^2+w_2 \z_4^2+w_6 \z_3\z_4+2w_5 \z_4)=R^{{\cal M}^5}
\la{crucialeq}
\eea
%%%
and these equations will determine the relations between $\a,\b,\g$.
Interestingly, it turns out that the four equations \rf{weq} are
automatically satisfied. This implies that one can freely choose $(a,b,c,d)$; as far as
the remaining equations in \rf{wconstr2},
%%%
\bea
&&  \b^2(w_3\z_1^2+w_2\z_2^2+w_6\z_1\z_2)=-\fr12 \g^2\nn\\
&& \z_1=-a,\quad \z_2=-b,\quad   \z_3=-a+c,\quad \z_4=-b+d \nn\\
&& \a^2(-1+2w_4 \z_3+w_3 \z_3^2+w_2 \z_4^2+w_6 \z_3\z_4+2w_5 \z_4)=R^{{\cal M}_5},
\la{remeq}\nn\\
\eea
%%%
are satisfied, the reduction will be consistent.
The first and third equation in \rf{remeq} becomes
%%%
\bea
\b^2&=&  \g^2\nn\\
\fr15 \a^2 &=& R^{{\cal M}_5}
\eea
%%%
regardless of values of $(a,b,c,d)$
, and therefore one must take
%%%
\bea
{{\cal M}_5}={S_5}
\eea
%%%
The case we have considered in the previous
subsection corresponds to\footnote{One can also consider the case
%%%
\bea
a=\fr83,\quad d=1,\quad b=c=0
\eea
%%%
This choice casts the exponential factors in \rf{fstotq2} into the forms that were considered
in \cite{Sato:2002kv}. One finds, in this case,
%%%
\bea
u_1=-4,\quad u_2=\fr52,\quad u_3=\fr{16}9,\quad u_4=\fr54,\quad u_5=-\fr{20}3
\eea
%%%
Some of these coefficients are different from those that appeared in \cite{Sato:2002kv}
, and should be an indication that \rf{ans1q0} and \rf{ans2q0} belong to a different class of ansatze than those of \cite{Sato:2002kv}.
Even though the two ansatze are different, they admit the same DBI solutions; we take this as certain robustness
of the DBI form solution.
}
%%%
\bea
a=d=1,\quad b=c=0
\eea
%%%
which leads to
%%%
\bea
u_1=-\fr32,\quad u_2=\fr52,\quad u_3=\fr14,\quad u_4=\fr54,\quad u_5=-\fr52
\eea
%%%
and
%%%
\bea
w_1=0,\quad w_2=-\fr3{10},\quad w_3=-\fr12,\quad w_4=-\fr12,\quad w_5=\fr12,\quad w_6=0
\eea
%%%

\subsection{Appearance of worldvolume gauge field }

The appearance of a gauge field through a spontaneous symmetry breaking should be a general phenomenon
independent of coefficients in $H$. Indeed, it is a general phenomenon as proved by
the following observation. The solution $S$ that appears in \rf{pianddelSq} can be viewed
as a functional of an antisymmetric "moduli field", $F_{\m\n}$,
%%%
\bea
S=S[F_{\m\n}]
%\qquad {\color{blue}(my~proposition~is)~ S=\sqrt{-g}f[F_{\m\n}]}
\eea
%%%
in the supergravity background.
To be able to view $F_{\m\n}$ as the field strength of a gauge field, the field equation and closure of $F_{\m\n}$ must be established. We examine the solutions of the non-constant/constant supergravity field case in this regard.    

{
Let us take the non-constant field solution of \rf{s01}, the higher order of which is expected to contain the WZ term of \rf{ScSBISWZ66} (cf. \cite{Shiromizu:2003dr}), and consider $Z^\m_C=0$ constraint in \rf{constraintsq} ($Z^\m_C$ is defined in \rf{HZconstr}),
%%%
\be
\N_\n (\fr{\g}{2\sqrt{-g}} \e^{\m\n\l\r}\mathcal{F}_{\l\r})+\fr{\g}{6\sqrt{-g}}\e^{\m\n\l\r}H_{\n\l\r}=0.
\ee
%%%
It follows that
\be
3\e^{\m\n\l\r}\N_\n \mathcal{F}_{\l\r}+\e^{\m\n\l\r}H_{\n\l\r}=0.
\ee
This equation implies, for $H_{\n\l\r}=3\pa_{[\n}B_{\l\r]}$,
%%%
\be
\mathcal{F}_{\l\r}=\pa_\l A_\r-B_{\l\r}
\ee
%%%
where $A_\r$ is the ``moduli" field that can be interpreted as the worldvolume gauge field.\footnote{
One subtlety is a question whether $F_{\m\n}$ would be abelian or
non-abelian. We comment on this and related issues in the conclusion.
}
 With $\mathcal{F}_{\m\n}=F_{\m\n}-B_{\m\n}$, the constraint $\N^\n \pi_{B \m\n}=0$ turns into
%%%
\be
\N_\m (e^{3/2 (\f+\r)}\mathcal{F}^{\m\n})+\dots=0.
\ee
%%%
where $(...)$ is an expression that contains, in particular, $\N_\m *F^{\m\n}$.
It is the field equation for the worldvolume gauge 
field $A_\m$, which comes from (higher-order extended) equation \rf{s01}.

The $\pi_C^{\m\n}$ also reveals information on the Hodge dual of $F_{\m\n}$. Let us take a covariant derivative on \rf{pianddelSq} with antisymmetrization,
%%%
\bea
\N_{[\k}\pi_C^{\mu\nu]}(x)=\N_{[\k}\frac{1}{\sqrt{-g(x)}}
\frac{\delta S}{\delta C_{\mu\nu]}(x)}=\g \fr14 \N_{[\k}\ast F^{\m\n]} \la{dpimn}
\eea
%%%
where $\ast F^{\m\n}$ is the Hodge dual field.
In the approximation of the constant supergravity fields at fixed $r$\footnote{This should be viewed as a leading order analysis in the derivative expansion.}, this yields the Bianchi identity for the $\ast F^{\m\n}$.
In the usual flat space case, the field equation and Bianchi identity are interchanged under Hodge duality. It will be interesting to explore this issue in the setup of the non-constant gravity fields.  
Note the interplay between $Z^\m_C$ and $Z^\m_B$ constraints: one of them has led to the Bianchi identity, and the other one to the $A_\r$ field equation. In the duality-symmetric approach to D3-brane \cite{Nurmagambetov:1998gp} these constraints are recast into a $SL(2,R)$ doublet, and lead to a single equation of motion of the duality-symmetric gauge field $A^i_\r$, $i=1,2$.

}

\subsection{Implications}

In the previous subsection, a gauge theory action was obtained after the HJ procedure. The HJ principal function $S$ is nothing but the lagrangian with $r$ playing the role of time. As mentioned in the introduction, the gauge action can be interpreted as a dual action to the 4D gravity action. The 4D gravity system itself is a (gauged) dS supergravity.

\subsubsection{on realization of "braneworld"}

Let us ponder whether the "braneworld" is realized by the current procedure.
First of all, we should note that the current procedure implies a qualitatively different
braneworld from the conventional Randall-Sundrum type in that the only dynamical degrees of freedom
are those of the gauge multiplet after integrating out the gravitational degrees of freedom.

The situation is analogous to the usual QFT procedure where instantons become
dynamical degrees of freedom that are "dual" to the original gauge theory. One has instanton
moduli and fluctuation degrees of freedom in the path integral once one expands around an
instanton solution. After one integrates out
the fluctuation degrees of freedom, one finds an instanton action that can be viewed
as "dual" to the original action (See \cite{Hashimoto:2005qh} and \cite{Hatefi:2012sy} for related discussions.)
The HJ procedure is a solution-finding procedure, and we saw the moduli field $F_{\m\n}$ enter for the case at hand. One should then integrate out the 4D gravitational degrees of freedom (i.e., all the other degrees of freedom than the moduli field), and eventually
find an action of the moduli field. There will also be the gravitational part of the background, therefore, the ultimate action of the moduli fields would be in that gravity background.\footnote{The action in the curved background might be identified as including quantum and none-perturbative effects. Such an identification was made
in \cite{GonzalezRey:1998uh} for example.}

For the braneworld realization, it would be required to check whether a brane solution
of \rf{szero} localizes at some value of $r$. As stated above, the current procedure leads to a qualitatively different braneworld.
 There still exists a feature within the current setup that might be an indication of the localization
of all the degrees of freedom:
%%%
\bea
\frac{\pa S}{\pa r_0}=0.
\eea
%%%

\subsubsection{new paradigm for black hole physics}

As discussed in \cite{Park:2009mi}, the black hole information paradox is an amenable problem in string theory
context. It is in the 4D pure Einstein gravity where the paradox becomes more subtle.
The present work may have an application in black hole physics; in particular,
in the aspect associated with the information paradox in the 4D pure Einstein gravity.

In the usual approach of QFT in a curved spacetime,
the geometry enters as a background whereas the matter fields are treated on the quantum level.
It is almost evident that geometry as a non-dynamic background would be inadequate
for describing physics in which the back-reaction plays a crucial role.
The information paradox should lie in the classical treatment of the geometry (see, e.g., the recent discussion in \cite{Brustein:2012jn}). The geometry is strictly classical in the conventional approach because
the matter quantum fields in the usual approach do not directly describe the fluctuations of the {\em geometry}.\footnote{It is the matter fields that represent the
fluctuating degrees of freedom in the conventional approach, and unlike the current ADM reduction approach, the matter fields are extrinsic to the geometric degrees of freedom. One may say that the matter fields indirectly describe the fluctuations of the geometry
since they are coupled to the metric. However, this would be so only when the geometry degrees of freedom are quantized as well.
 } The best solution for
this status of matter would be the full quantization of the Einstein-Hilbert action. Given the unavailability
of such an apparatus, the second best solution would be to have
a {\em semi-classical} treatment of geometry.
The dual gauge action obtained through the ADM reduction of this paper should provide the needed semi-classical tool.

In the usual approach, the matter fields are not
intrinsically gravitational degrees of freedom. In contrast, the gauge action obtained
as a result of ADM reduction provides degrees of freedom that are intrinsic to the original
gravity system.
 The matter field equations are solved in some background metric in the conventional
approach. Since
it is not the proper full coupled equations between matter and metric that are solved, the result is bound to be without a proper account of back-reaction from the metric
that gets deformed by the matter.
In the proposed ADM reduction approach, one gets the
"matter" system, i.e., the YM field after the spontaneous symmetry breaking. In other words, the appearance of the "matter fields" is built into the formulation.
One can then try to solve those equations associated with YM field. However, the interpretation is now very different: the gauge field equations directly, although semi-classically, describe the fluctuations of the geometry. We will have more on this as well as other speculative issues in the conclusion.

%%%%%%%%%%%%%%%%%%%%%%%%%%%%%%%%%%%%%%%%%%%%%%%%%%%%%%
\section{Domain-wall solution, toroidal compactification and "inflaton"}

\renewcommand{\theequation}{4.\arabic{equation}}
\setcounter{equation}{0}

In this section, we analyze two more aspects of the 5D action \rf{5DLagEins} that has been obtained by
the sphere reduction in sec 2.
One thing to note is that even if we are using the notation $r$, it is not necessarily a radial coordinate;
it is one of the spatial coordinates.

\subsection{Domain-wall solution}

One may use either a 5D Einstein-type frame or a "string"-type frame to find a solution. In this section, we use an Einstein type frame. (It should
be possible to find the corresponding solution in a 5D "string-type" frame.) Consider \rf{5dactRhoEinq} which we quote here for convenience,
%%%
\bea
I_5
   &=& \fr1{2\k_5^2}\int d^5\xi \sqrt{-h}\Big[ R^\5-\fr56(\pa\r)^2
-\fr12(\pa \f)^2-\fr1{2\cdot 3!}  e^{-\phi}e^{\fr53 \r} H_\3^2\nn\\
    &&  -\fr12 e^{2\phi}(\pa\chi)^2-\fr1{2\cdot 3!}e^{\phi}e^{\fr53 \r}\Ft^{2}_\3-\fr1{2\cdot 5!}e^{\fr{10}3 \r}\Gt^2_\5+e^{-\fr43\r} R^{{\cal M}_5}\Big]
    \label{5dactRhoEinqq}
\eea
%%%
and the reduced field equations setting $\chi=H=\f=F=0$. In this section, we take
%%%
\bea
R^{{\cal M}_5}= R^{S^5}
\eea
%%%
Below we will set $G=0$ as well because only that case admits a relatively simple
solution.
It follows from the action given in \rf{5dactRhoEinq} with $F=\chi=H=\f=0$ that
%%%
\be
\N_{\um}\left(e^{\fr{10}{3}  \r}\, \Gt_\5^{\;\;\um \un_1...\un_4}\right)=0,
\la{Gtfeq}
\ee
%%%
%%%
\be
\N^2\r-\fr1{120}e^{\fr{10}3\r}\Gt^2_\5-\fr4{5} e^{-\fr43\r} R^{{\cal M}_5}=0
\la{rhofeq}
\ee
%%%
%%%
\bea
R^\5_{\um\un} -\fr5{6}\pa_{\um} \r \,\pa_{\un} \r
-\fr{ 1}{4\cdot 4!}e^{\fr{10}3 \r}\, \Gt_{\um\up\uq\ur\us}{\Gt_{\un}}^{~\up\uq\ur\us}-h_{\um\un}\left(\fr12 R^\5 -\fr{5}{12}(\pa \r)^2+\fr12 e^{-\fr43\r}R^{{\cal M}_5} \right)=0\nn\\
\la{Einsfeq}
\eea
%%%
Let us try the following metric ansatz,
%%%
\bea
ds_5^2= e^{2A}dr^2+e^{2C(r)}ds_{dS_4}^2
\la{ds5sol2q}
\eea
%%%
The $G_5$ field equation \rf{Gtfeq} implies
%%%
\bea
G_5^{m_1...m_5} &=& \fr{k}{\sqrt{-h}}e^{-\fr{10}{3}\r}\e^{m_1...m_5} \nn\\
G_5^2 &=& -5!k^2 \;e^{-\fr{20}{3}\r}
\eea
%%%
\be
\Gt_{\um \un_1 \dots \un_4}{\Gt_{\uk}}^{\,\,\,\un_1\dots \un_4}=\fr15 h_{\um\uk}\Gt^2_\5 \,.
\la{Gtrel5q}
\ee
%%%
Consider $(rr)$ and $(11)$ components of \rf{Einsfeq}:
%%%
\be
R^\5 -\fr{5}{6}h^{rr}(\pa_r \r)(\pa_r \r)+ e^{-\fr43\r}R^{{\cal M}_5}
-\fr2{h_{rr}}\Big[R^\5_{rr} -\fr{5}{6}(\pa_r \r)(\pa_r \r)-\fr{1}{4\cdot 4!} e^{\fr{10}3 \r}\Gt_{r\up\uq\ur\us}{\Gt_{r}}^{~\up\uq\ur\us}\Big]=0
\la{Einsrr}
\ee
%%%
%%%
\be
R^\5 -\fr{5}{6}h^{rr}(\pa_r \r)(\pa_r \r)+ e^{-\fr43\r}R^{{\cal M}_5}
-\fr2{h_{11}}\Big[R^\5_{11} -\fr{1}{4\cdot 4!} e^{\fr{10}3 \r} \Gt_{1\up\uq\ur\us}{\Gt_{1}}^{~\up\uq\ur\us}\Big]=0
\la{Eins11}
\ee
%%%
Combining \rf{Einsrr} and \rf{Eins11}, one gets
%%%
\bea
\fr1{h_{rr}}\Big[R^\5_{rr} -\fr5{6}\pa_{r} \r \,\pa_{r} \r \Big]
=\fr1{h_{11}}R^\5_{11}  \la{difone}
\eea
%%%
Using the result, e.g., in appendix B of \cite{Douglas:2009zn}, one can show
%%%
\bea
R_{11}^\5&=& R_{11}^\4+g_{11}e^{2C-2A}(-4 \N_r C\N_r  C-\N_r \N_r C
  +\N_r A \N_r C)  \nn\\
R_{rr}^\5&=& 4(\N_r A\N_r C-\N_r C \N_r C-\N_r\N_r C)
\eea
%%%
With these, \rf{difone} yields\footnote{
where
%%%
\bea
\N_r \N_r= \N^2 =e^{-A}\pa_r (e^{-A}\pa_r)
\eea
%%%
}
%%%
\bea
&& 4\N_r A \N_r C- 4\N_r C \N_r C-4\N_r  \N_r C -\fr56 (\pa_r\r)^2  \nn\\
=&&e^{2A-2C}(\L -4 e^{2C-2A}\N_r C\N_r  C-e^{2C-2A}\N_r \N_r C
  +e^{2C-2A}\N_r A \N_r C )    \nn\\
 \la{difone2}
\eea
%%%
Let us take $h^{mn}$ on \rf{Einsfeq}:
%%%
\bea
R^\5 -\fr{5}{6}h^{rr}(\pa_r \r)(\pa_r \r)+\fr53 e^{-\fr43\r}R^{{\cal M}_5}
 +\fr{1}{6\cdot 4!} e^{\fr{10}3\r}G_5^2=0  \la{EinsRicci}
\eea
%%%
Combining \rf{EinsRicci} and \rf{Eins11} and setting $G=0$\footnote{If one keeps $G$, two different types of $e^{\r}$ factors are present, and this eliminates possibility of any simple solution.}, one gets
%%%
\bea
-\fr23 e^{-\fr43 \r}R^{{\cal M}_5}=2 e^{-2C}(\L^\4-4e^{2C}\N_r C \N_r C-e^{2C}\N_r\N_r C)
\la{ricci11}
\eea
%%%
Then \rf{ricci11} implies
%%%
\bea
C=\fr23 \r\quad,\quad \L^\4=3q_2^2-\fr13 R^{{\cal M}_5}
\eea
%%%
The $\r$-eq \rf{rhofeq} takes
%%%
\bea
\fr{d^2 \r}{dr^2}+(4 \fr{dC}{dr}-\fr{dA}{dr})\fr{d\r}{dr}-\fr45 e^{2A-\fr43\r}R^{{\cal M}_5}=0
\la{rhoeq4}
\eea
%%%
In the absence of $G_\5$, one can show that \rf{Eins11} takes
%%%
\bea
&& R^\5-\fr56 e^{-2A}(\fr{d\r}{dr})^2+ e^{-\fr43 \r}R^{{\cal M}_5}=\fr{2}{h_{11}}R_{11}^\5
\la{Eins112}
\eea
%%%
Let us consider the following set of ansatze:
%%%
\bea
&& A=0\nn\\
&&\r=p\ln(q_1+q_2\, r)\nn\\
&&C=\fr23 p\ln(q_1+q_2\, r)  \la{rhoc}
\eea
%%%
Using
%%%
\bea
&& R_\5= e^{-2C}R_\4+4\Big[
      -2 \N_r\N_r C-5 \N_r C \N_r C\Big]\nn\\
\eea
%%%
\rf{Eins112} takes
%%%
\bea
&& e^{-2C}R^\4-4(2\N_r\N_r C+5\N_r C\N_r C)+e^{-\fr43 \r}R^{{\cal M}_5}-\fr56 e^{-2C}(\pa_r \r)^2\nn\\
=&&  2e^{-2C}(\L-4 e^{2C}\N_r C\N_r C- e^{2C}\N_r \N_r C)  \la{R11eq}
\eea
%%%
Substituting \rf{rhoc} into \rf{rhoeq4} and \rf{difone2}, one can see that $p=\fr32$
and
%%%
\bea
&& R^{{\cal M}_5}=\fr{45}{8}q_2^2\quad \mbox{from \rf{rhoeq4}} \nn\\
&& \L=\fr{9}{8}q_2^2  \quad \mbox{from \rf{difone2}}
\eea
%%%
where $\L$ ($\equiv \L^\4$) is defined by $R_{\m\n}^\4=\L g_{\m\n}$. Eq.\rf{R11eq} also produces a consistent result:
%%%
\bea
&& 2\L+R^{{\cal M}_5}=\fr{63}{8}q_2^2
\eea
%%%

\subsection{Toroidal compactification and "inflaton"}

Carrying out a conventional dimensional reduction on
the 5D action \rf{5dactRhoEinq} will be worthwhile because it will yield
a 4D gravity theory with various gauge fields with positive cosmological constant.
 The theory has been obtained from IIB supergravity, and provides a potentially interesting inflationary model. Let us consider the $\r$-rescaled form \rf{5DLagEins1}. One can easily carry out dimensional reduction keeping as many fields
in \rf{5DLagEins1} as one wished. Focusing on the perspective of 4D inflatonary physics, we illustrate
the case with the metric and $\r$:
%%%
\bea
I&=& \fr1{2\k_5^2}\int d^5\xi \sqrt{-h}\Big[ R^\5-\fr12(\pa\r)^2
+e^{-\fr{4}{\sqrt{15}}\r} R^{{\cal M}^5}\Big]
    \label{5dactRhoEinqred}
\eea
%%%
One can consider a simple dimensional reduction given by
%%%
\bea
\r=\r(x^\m)\quad,\quad ds_5^2=dr^2+g_{\m\n}dx^\m dx^\n
\eea
%%%
The resulting 4D action takes
%%%
\bea
I&=& \int d^4 x \sqrt{-g}\Big[ R^\4-\fr12(\pa\r)^2
+e^{-\fr{4}{\sqrt{15}}\r} R^{{\cal M}^5}\Big]
    \label{4dinf}
\eea
%%%
 The gauge fields $\chi ,C, D$ and $B$ can easily be accommodated.
 It would be interesting to investigate whether one could construct a dS solution with
 addition of various form fields. If so, that would be in line with the observation made in
 the KKLT type approaches.
 We leave the resulting model's phenomenological study for the future.

%%%%%%%%%%%%%%%%%%%%%%%%%%%%%%%%%
\section{Conclusion}
%%%%%%%%%%%%%%%%%%%%%%%%%%%%%%%%%

In sec 2, we have carried out reduction of IIB supergravity on ${\cal M}_5 (=\cH_5\;\; \mbox{or}\;\; S_5 )$. The resulting 5D gravity lagrangian has been analyzed in
several different directions. In one direction, we have shown that it admits a 4D curved domain-wall solution.
In another direction, we have performed, following \cite{Sato:2002kv,Sato:2003ky,Sato:2004ic}, the Hamilton-Jacobi procedure of canonical transformation and have obtained another gravity description.
As shown in sec 3, the Hamilton-Jacobi equation admits a class of solutions that take a form of a gauge theory action.\footnote{The fact that the original lagrangian admits a domain-wall solution and its Hamilton-Jacobi equation admits a worldvolume action as a solution should be related although we will not pursue this aspect on a deeper level than is apparent.}

The way the gauge field strength $F_{\m\n}$ appears is intriguing. The worldvolume gauge fields emerge as "moduli fields":
regardless of the values that the gauge fields take, the gauge action satisfies
 the Hamilton-Jacobi equation of the gravity system, and the gauge fields describe the fluctuations of the moduli space. They must be an inequivalent set of extremum solutions, and the inequivalence
must stem from different patterns of the brane fluctuations. Those patterns are parameterized through
the field $F_{\m\n}$. Interestingly, the appearance of the dual degrees of freedom as a form of moduli fields
was observed before in the context of forward duality: the strong coupling limit of
a DBI action admits a class of solutions that can be collectively interpreted as a closed string
action \cite{Nielsen:1973qs,Gibbons:2000hf,Sen:2000kd,Park:2001bm}.

One of the potentially powerful implications is the fact that getting
non-gravitational degrees of freedom through ADM reduction would work for the pure 4D
Einstein-Hilbert action. We believe that for the proper treatment of the black hole information
two conditions are required for the QFT tool adopted to tackle the paradox. The first is that the adopted QFT should {\em directly} describe the fluctuations of the geometry.
It is necessary to use a formulation that is self-consistent or "closed" under the forward and backward dualizations.
 The second is that the QFT interactions in their precise forms must be included.
The ADM reduction would, in principle at least, determine the precise form of the interactions of the resulting gauge theory. In these regards, the ADM reduction approach should provide a proper paradigm for black hole physics.

\vspace{.3cm}

There are multiple future directions:

\vspace{.1cm}

\ni One is an obvious direction of studying the supersymmetry aspect of the 5D/4D theories.
Several other directions are associated with a better understanding of the ADM reduction itself.
One may try to extend the program of sec 3.1 to higher orders in the derivative expansion.
Another direction would be to address the following question: what at the full string theory level would be responsible for the appearance of a gauge field from a gravity system? The appearance of a gauge field
seems to be a general phenomenon that occurs in a low energy theory that
may not have embedding in a string theory. However, it would be still interesting to see
the full stringy mechanism that is behind for the theories that do have stringy embedding.
(See our speculation below.)

There are other related issues that require further study. 
In section 3, we have scratched the surface of the phenomenon of the gauge field emergence.
We will present a more thorough and comprehensive analysis elsewhere. 
Related is the issue of whether the emerging gauge field would be abelian or non-abelian.
 The appearance of an abelian gauge moduli field is straightforward. The real
 question is if there could be a (relatively simple) way to introduce non-abelian degrees of freedom. At this point we can only state what we anticipate and should postpone a better answer until further research. Presumably, the abelian vs non-abelian issue
 would depend on whether one uses a collection of D3 branes or, say D1 branes to describe the bulk physics. As observed, e.g., in \cite{Nurmagambetov:2011pu}, a higher dimensional
abelian brane can be described by a non-abelian lower dimensional branes.

\vspace{.1cm}

As stated in sec 2, applications of the ADM reduction approach to black hole information should be interesting as well.
The other directions concern phenomenological aspects and applications.
One may take \rf{4dinf} with other form fields as a starting point, and study the resulting
  4D Friedmann-Lemaitre-Robertson-Walker
 eqs in the presence of D3 (or even D7).
It will be interesting to make a connection this way with the KKLT and related compactification
scenarios.

\vspace{.2cm}

Finally remarks on more speculative aspects are in order:
The {\em forward dualization} mentioned in the introduction
should be associated with endpoints of an open string sticking together and becoming a closed string. By the same token, the appearance of gauge degrees of freedom should presumably be associated with a closed string opens up and becomes an open string on the closed string theory level. It would be very interesting if one could make this more precise and quantitative.

The following question was raised in \cite{Sato:2003uc}. The DBI action contains all $\a'$-order terms\footnote{Of course, this is true only in the leading derivative expansion in $\F_{\m\n}$; once the subleading terms such as $\pa \F$, $\pa\pa \F$, etc. are taken into account, new terms would appear.} but it still appears as a solution of reduced supergravity that is just the leading $\a'$ action of a closed string.
Perhaps the answer lies in the following. The gauge form solution represents excitations of
massless open string modes. The higher $\a'$ corrections to the IIB supergravity may be associated with a massive gauge action that has all the {\em massive open string modes} appearing explicitly at first and then subsequently integrated out in the open string context, therefore, deforming the massless gauge field. (The integrating out procedure should be done using the full string theory setup which of course would be a hard step in practice.) Differently put, the massless closed strings viewed as a composite open string state should be massive, and apparently they seem sufficient to
account for the massless gauge theory modes.

\ssk
\ssk
{\bf Acknowledgements.} AJN acknowledges partial support from the Joint DFFD-RFBR Grant \# F40.2/040.
Part of this work was carried out during IP's visit to CQUeST, Sogang university.
IP thanks B. H. Lee, J. H. Park and C. Rim for their hospitality.

\newpage
\renewcommand{\theequation}{A.\arabic{equation}}
 \setcounter{equation}{0}
  \section*{Appendix A: Differential form conventions  }

We use the following conventions on differential forms.
The flat space metric signature is mostly positive, so $\det \eta_{ab}=-1$ in any space-time dimensions. The Levi-Civita tensor $\e^{a_1\dots a_D}$ is defined such that
%%%
\be
\e^{01\dots D-1}=1,\quad \e_{01\dots D-1}=-1,
\la{LCD}
\ee
%%%
hence
\be
\e^{a_1\dots a_D}\e_{a_1\dots a_D}=\e_{a_1\dots a_D}\eta^{a_1a^\pr_1}\dots \eta^{a_D a^\pr_D}\e_{a^\pr_1\dots a^\pr_D}=\det \eta \cdot D!=-D!
\la{LCLCD}
\ee
Generalization of \rf{LCLCD} to a curve background is given by
%%%
\be
\e^{m_1 \dots m_D}\e_{m_1 \dots m_D}=\det g \cdot D!
\la{LCLCcD}
\ee
%%%
For a $p$-form we choose
\be
\Omega_{(p)}=\fr1{p!}\Omega_{m_1\dots m_p} \, dx^{m_1}\wg \dots \wg dx^{m_p},
\la{pform}
\ee
and
\[
d\Omega_{(p)}=\fr1{p!}\pa_{[m_{p+1}}\Omega_{m_1\dots m_p]}\, dx^{m_{p+1}}\wg dx^{m_1}\wg \dots dx^{m_p}
\]
\be
=\fr1{p!}\pa_{[m_1}\Omega_{m_2\dots m_{p+1}]}\, dx^{m_1}\wg dx^{m_2}\wg \dots dx^{m_{p+1}}.
\la{dpform}
\ee
%%%
%%%
The external derivative $d$ acts from the left, i.e.,
%%%
\be
d(\Omega_{(p)} \wg \Omega_{(q)})=d\Omega_{(p)}\wg \Omega_{(q)}+(-)^p \Omega_{(p)} \wg d\Omega_{(q)} .
\la{dpq}
\ee
%%%
We define the Hodge star as
\be
\ast \left(dx^{n_1}\wg \dots dx^{n_p} \right)=\fr1{(D-p)!} \fr1{\sqrt{-g}} \, {\e_{{m_1}\dots m_{D-p}}}^{n_1 \dots n_p} \, dx^{m_1}\wg \dots dx^{m_{D-p}},
\la{astD}
\ee
so the dual to $\Omega_{(p)}$ form
$ \ast \Omega_{(p)}$  is defined by
\be
%\tilde{\Omega}_{(D-p)}
\ast{\Omega}_{(p)}=\fr1{(D-p)!p!}\fr1{\sqrt{-g}}\, {\e_{{m_1}\dots m_{D-p}}}^{n_1 \dots n_p} \, \Omega_{n_1\dots n_p} \, dx^{m_1}\wg \dots dx^{m_{D-p}}.
\la{dualpD}
\ee
On account of the latter expression and eq. \rf{LCLCcD}, one gets
%%%
\be
\ast^2_p=(-)^{Dp+p+1}.
\la{ast2}
\ee
%%%
To get \rf{ast2}, the following relation should be used:
%%%
\be
{\e_{{m_1}\dots m_{D-p}}}^{n_1 \dots n_p} {\e^{{m_1}\dots m_{D-p}}}_{k_1 \dots k_p}=\det g \cdot (D-p)! p! \d^{[n_1}_{[k_1} \dots \d^{n_p]}_{k_p]}\equiv
\det g \cdot (D-p)! p!  \d^{n_1 \dots n_p}_{k_1 \dots k_p} \,.
\la{eepD}
\ee
%%%
Also, we have
\be
dx^{m_1}\wg \dots \wg dx^{m_D}=d^Dx \, \e^{m_1 \dots m_D},
\la{dxVD}
\ee
Taking into account \rf{pform}, \rf{dualpD} and \rf{eepD}, this implies
%%%
\be
\fr1{2 \cdot p!}  \int \, d^D x \, \sqrt{-g} \left( F_{(p)} \right)^2=(-)^{Dp+p+1}\, \fr12 \int_{\mathcal{M}^D} \, F_{(p)} \wg \ast F_{(p)} .
\la{Fp2D}
\ee
%%%
In our notation
%%%
\be
\int \, d^D x \, \sqrt{-g} \equiv \int_{\mathcal{M}^D} \, \mathbf{1}  \, ,
\la{VD}
\ee
%%%
and, therefore,
%%%
\be
\int \, d^D x \, \sqrt{-g} R \equiv \int_{\mathcal{M}^D} \, \mathbf{1} \cdot R \, .
\la{EHdiffD}
\ee

\newpage
%%%%%%%%%%%%%%%%%%%%%%%%%%%%%%%%%%%%%%%%%%%%%%%%%%%%%%%%%%%%%%%%

\end{document}